\newcommand{\version}{November 30, 2018}
\documentclass[letterpaper,oneside,11pt,onecolumn,pdftex]{article}
\pdfoutput=1
\usepackage[bindingoffset=0.0cm,textheight=22.6cm,hdivide={*,16.0cm,*}, vdivide={*,22.6cm,*}]{geometry}
\usepackage{amsmath,amsfonts,amssymb}
\usepackage{mathtools}
\usepackage[pdftex]{graphicx}
\usepackage[margin=1cm,font=small,labelfont=bf]{caption}
\usepackage[T1]{fontenc}
\usepackage[utf8]{inputenc}
\usepackage{fouriernc}
\newcommand{\id}{\mathbb{1}}

 \usepackage[numbers,square,comma,sort&compress]{natbib}

\usepackage[pdftex,hyperref,svgnames]{xcolor}
\usepackage[pdftex,bookmarksnumbered=true,breaklinks=true,%
colorlinks=true,linktocpage=true,linkcolor=MediumBlue,citecolor=ForestGreen,urlcolor=DarkRed]{hyperref}

\makeatletter
\AtBeginDocument{
  \hypersetup{
    pdfsubject = {\Abstract},
    pdfkeywords = {\keywords},
    pdftitle = {\@title},
    pdfauthor = {\@author}
  }
}
\makeatother

\makeatletter

 \@addtoreset{equation}{section}
 \makeatother

\renewcommand{\d}{\delta}
\newcommand{\e}{\epsilon}

\renewcommand{\th}{\theta}
\newcommand{\vth}{\vartheta}

\renewcommand{\l}{\lambda}
\newcommand{\m}{\mu}

\newcommand{\rh}{\rho}
\newcommand{\s}{\sigma}

\newcommand{\ph}{\phi}

\newcommand{\w}{\omega}

\newcommand{\G}{\Gamma}

\newcommand{\Th}{\Theta}

\newcommand{\W}{\Omega}

\newcommand{\ml}{\mathfrak{l}}
\newcommand{\mm}{\mathfrak{m}}
\newcommand{\mn}{\mathfrak{n}}

\newcommand{\pa}{\partial}
\newcommand{\inv}[1]{\frac{1}{#1}}

\newcommand{\nn}{\nonumber}
\newcommand{\eqnref}[1]{Eq. \eqref{#1}}

\newcommand{\co}[2]{\left[#1,#2\right]}

\newcommand{\bfa}{\mathbf{a}}
\newcommand{\bfw}{\mathbf{w}}

\newcommand{\kb}{k_{\txt{B}}}

\newcommand{\ct}{c_{\textrm{t}}}
\newcommand{\cl}{c_{\textrm{l}}}

\newcommand{\bt}{\beta_{\textrm{t}}}

\newcommand{\wl}{\w_{\textrm{l}}}
\newcommand{\wt}{\w_{\textrm{t}}}
\newcommand{\Vc}{V_\txt{c}}

\newcommand{\te}{\tilde{d}}
\newcommand{\elA}{\tilde{A}}

\newcommand*{\vv}[1]{\vec{\mkern0mu#1}}
\newcommand{\vq}{\vv{q}}
\newcommand{\vqi}{\vv{q}\,^{\!'}}
\newcommand{\vqii}{\vv{q}\,^{\!''}}
\newcommand{\qt}{\tilde{q}}
\newcommand*{\mat}[1]{\mathrm{#1}}

\newcommand{\txt}[1]{\textrm{#1}}

\DeclarePairedDelimiter\abs{\lvert}{\rvert}
\newcommand{\coleq}{\vcentcolon=}

\title{\texorpdfstring{\begin{flushright}
        {\small LA-UR-18-22746}
       \end{flushright}\vspace{2em}}{}%
       Velocity dependent dislocation drag from phonon wind \texorpdfstring{\\}{}%
       and crystal geometry}

\author{Daniel N. Blaschke}

\date{\version}

\newcommand{\Abstract}{%
The mobility of dislocations is an important factor in understanding material strength.
Dislocations experience a drag due to their interaction with the crystal structure, the dominating contribution at high stress and temperature being the scattering off phonons due to phonon wind.
Yet, the velocity dependence of this effect has eluded a good theoretical understanding.
In a previous paper, dislocation drag from phonon wind as a function of velocity was computed from first principles in the isotropic limit, in part for simplicity, but also arguing that macroscopically, a polycrystalline metal looks isotropic.
However, since the single crystal grains are typically a few microns up to a millimeter in size, dislocations travel in single crystals and cross boundaries, but never actually see an isotropic material.
In this work we therefore highlight the effect of crystal anisotropy on dislocation drag
by accounting for the crystal and slip plane geometries.
In particular, we keep the phonon spectrum isotropic for simplicity, but dislocations are modeled according to the crystal symmetry (bcc, fcc, hcp, etc.).
We then compare to the earlier purely isotropic results,
as well as to experimental data and MD simulations where they are available.
}

\newcommand{\keywords}{dislocations in crystals, drag coefficient, phonon wind}

\graphicspath{{./}{./figures/}{./figures/disloc_core/}}

\newcommand{\pdfinsert}[2]{
\includegraphics[width=\textwidth]{#1.pdf}
\caption{#2}
}

\begin{document}

\maketitle

\thispagestyle{empty}
\begin{center}
\vspace{-0.3cm}
Los Alamos National Laboratory\\Los Alamos, NM, 87545, USA
\\[0.5cm]
\ttfamily{E-mail: dblaschke@lanl.gov}
\end{center}

\vspace{1.5em}

\begin{abstract}
\Abstract
\end{abstract}

\newpage
\tableofcontents

\section{Introduction}
\label{sec:intro}

A fundamental problem in the dynamic response of solid metals are the mechanisms contributing to the so-called drag coefficient of dislocations under high stresses and strains:
Moving dislocations (curvilinear defects in the crystal structure of the metal) experience a drag due to their interaction with the crystal structure, and represent a major factor in the understanding of material strength.
Hence, many dislocation based material strength models require the dislocation drag coefficient $B$ as one of their input parameters (typically determining the dislocation glide time between obstacles), see e.g.~\cite{Krasnikov:2010,Barton:2011,Hansen:2013,Hunter:2015,Borodin:2015,Luscher:2016,Austin:2018}.
$B$ is usually assumed to be a constant (or a constant over a simple ``relativistic'' factor) as a fist order approximation.
Hence, more insight into the true functional form of this drag coefficient could improve those models.

Several mechanisms contribute to the dislocation drag, and depending on the temperature, pressure and dislocation-velocity (or stress) regime, different mechanisms dominate~\cite{Nadgornyi:1988,Alshits:1992}.
For example, at low stresses, the dislocation mobility is limited by 
various potential barriers within the crystal.
Such obstacles can be overcome by a dislocation either by thermal activation (if the temperature is high enough) or by high enough stress levels.
When the stress level becomes ``critical'', i.e. high enough to easily overcome the highest potential barrier, the dislocation drag becomes viscous in character, and a significant change in the stress-velocity dependence from non-linear to approximately linear takes place.
In this high stress regime, where typical dislocation speeds are within a few percent of transverse sound speed, the dominating contribution to the dislocation drag coefficient (at temperatures around and above the Debye temperature) is the dissipative effect\footnote{Other dissipative effects, which we do not touch upon in this paper as they are subleading in the regimes we are interested in, are the so-called thermoelastic damping, the flutter effect, and the radiation damping, see Ref.~\cite{Nadgornyi:1988} for details.} of scattering off phonons (``phonon wind'').

The theory of phonon wind has a long history, being pioneered by Leibfried and others~\cite{Leibfried:1950,Lothe:1960,Nabarro:1961,Eshelby:1962,Brailsford:1972}, significantly improved from first principles by Alshits and collaborators~\cite{Alshits:1969b,Alshits:1973,Alshits:1979}, and was nicely reviewed in~\cite{Nadgornyi:1988,Alshits:1992} (which may also be consulted for additional references).
Due to the simplicity of Leibfrieds expression for dislocation drag ($B\sim$const.$\times T$), which represents the limit of high temperature $T$ and small dislocation velocity in an isotropic continuum, it is still used today (despite its limitations) as an empirical fitting function to extract information on dislocation mobility from discrete lattice simulations~\cite{Olmsted:2005,Marian:2006,Cho:2017}.
In these examples, the additional damping in the high velocity regime is then accounted for empirically by adding a $T$-independent term which grows like $\sqrt{v}$ above some threshold velocity $v>v_0$,
and which is based on Eshelby's arguments~\cite{Eshelby:1956} for screw dislocations in an isotropic continuum supplemented by an anisotropic dispersion relation.
The latter term is in stark contrast to the ``relativistic'' factors $\propto 1/(1-v^2/c^2)^m$ with different exponents $m$ and a limiting (sound) speed $c$ introduced by many authors (see e.g. Refs.~\cite{Krasnikov:2010,Barton:2011,Luscher:2016,Austin:2018} among others) based on equally empirical arguments.
Thus, a better understanding of dislocation drag from first principles at high velocities and for arbitrary crystal geometries is clearly needed.

For a wide range of velocities (already starting at low velocities where phonon wind is a subleading effect), the contribution to the drag coefficient due to phonon wind is roughly constant.
However, at very high velocities (i.e. more than a few percent of sound speed) the drag coefficient due to phonon wind becomes velocity dependent, indicating once more a non-linear stress-velocity dependence, and it is this regime we are primarily interested in here.

Existing continuum models of dislocation drag due to phonon wind~\cite{Alshits:1992} assume that the dislocation velocity is much smaller than the speed of sound in the material, and do well in describing the viscous regime.
However, for materials under high stress this assumption must be re-examined for a more realistic calculation of the dislocation drag coefficient, including its velocity dependence.
As a first step we study the velocity dependence in the subsonic regime, and intend to extend the theory to include dislocations moving at transonic and supersonic speeds in future work.
The motivation for the latter comes from recent MD simulations and experiments which indicate the existence of dislocations moving at supersonic speeds --- at least in certain materials such as plasma crystals~\cite{Nosenko:2007}, see also~\cite{Rosakis:2001,Li:2002,Olmsted:2005,Jin:2008,Pellegrini:2010,Gilbert:2011,Pellegrini:2014,Ruestes:2015} and references therein.

In a previous paper~\cite{Blaschke:BpaperRpt}, dislocation drag from phonon wind (from purely transverse phonons) was computed in the isotropic limit, mainly for simplicity, thereby generalizing the earlier models described in~\cite{Alshits:1992} to higher velocities.
However, since the single crystal grains are typically a few microns up to a millimeter in size, dislocations travel in single crystals and occasionally cross boundaries, but never actually see an isotropic material.
The purpose of the present paper is therefore to highlight the effect of crystal anisotropy on the dislocation drag coefficient from phonon wind by accounting for the crystal and slip plane geometries.
As a first step towards a more sophisticated model, we keep the phonon spectrum isotropic for simplicity, but dislocations are modeled according to the crystal symmetry (bcc, fcc, hcp, etc.).
We then compare to the purely isotropic results (now including also longitudinal phonons and thus generalizing~\cite{Blaschke:BpaperRpt}), seeing some deviations especially at high velocity, but even at small velocities for some materials.
For the isotropic limit we use experimental polycrystalline elastic constant data.
These deviations are expected since the present ``semi-isotropic'' approximation is able to capture features which are lost in the purely isotropic limit, such as the dislocation character dependence.
Additionally the uncertainties in the experimental determination of elastic constants (both single and polycrystalline) --- especially at third order --- might also contribute to the deviations seen between the two methods for pure screw and edge dislocations at low velocity.
In the high velocity regime, the observed large deviations between the two methods are expected since the position of divergences in the dislocation displacement gradient fields depends on the crystal geometry~\cite{Blaschke:2017lten}.

The outline of this paper is as follows:
In Section~\ref{sec:phononwind} we start by reviewing the phonon wind contribution to the drag coefficient in the continuum approximation, following Ref.~\cite{Blaschke:BpaperRpt} for the purely transverse phonons, and subsequently generalizing to include also longitudinal phonons in Section~\ref{sec:longitudinal}.
We then explain how to generalize the model to include anisotropic crystals, albeit assuming for simplicity an isotropic phonon spectrum.
In Section~\ref{sec:dislocations} we then review the method of deriving the displacement gradient field of a dislocation moving at constant (sub-sonic) velocity and define the slip systems considered in the present case; see~\cite{Blaschke:2017lten,Bacon:1980} and references therein for details.
Finally, in Section~\ref{sec:results} we present our results for dislocation drag in various metals of cubic, hexagonal, and tetragonal symmetry, and compare them to earlier experimental data, MD simulations, as well as our previous (more crude) purely isotropic model of~\cite{Blaschke:BpaperRpt},
albeit now including also longitudinal phonons.

\section{The phonon wind contribution to the drag coefficient}
\label{sec:phononwind}
\subsection{General considerations}

In this work, we consider the harmonic approximation (where displacements are small compared to the lattice spacings) and take the continuum limit.
We are interested in the interaction of phonons with a single moving dislocation in a crystal.
Details of the derivation of the according Hamiltonian in the continuum description can be found in Ref.~\cite{Blaschke:BpaperRpt} --- see also~\cite{Alshits:1992} and references therein for earlier work on this theory.
Hence our starting point is the following Hamiltonian\footnote{
Essential steps in deriving this expression are briefly outlined in Appendix~\ref{sec:appendix}, albeit we refer the interested reader to Refs.~\cite{Blaschke:BpaperRpt,Alshits:1992} for further details.
}:
\begin{align}
H&=H_0+H'(t) \,, \qquad\qquad
H_0=\sum_{\vq} \hbar\w_{q}\left(\bfa^\dagger_{q}\,\bfa_{q}+\frac12\right)
\,,\nn\\
 H'(t)&=\int_0^{2q_\txt{BZ}}\!\frac{dq \,q}{(2\pi)^2}\int_0^{2\pi}\!d\phi e^{-iqv\abs{\cos\phi}\,t}\sum_{\vqi}\G_{q',q'-q}(q,\phi)\xi_{q'}^\dagger \xi_{q'-q}
 \,, \nn\\
 \G_{q'q''}(q,\phi)&= \frac{\hbar}{4\rh \sqrt{\w_{q'}\w_{q''}}} \sum_{i,j,k}d_{kk'}(q,\ph) \bfw_{q'i} ^*\bfw_{q''j} 
 \sum_{i'j'k'} q'_{i'}q''_{j'} \elA_{ijk}^{i'j'k'}
 \,, \label{eq:Hamiltonian-start}
\end{align}
consisting of the usual kinetic part for the phonons $H_0$ and the interaction between phonons and the dislocation $H'$.
Following Ref.~\cite{Alshits:1992} we used the shorthand notation (or super-indices) $q'\coleq\{\vqi,s'\}$;
hence $\G_{q'q''}\coleq \G_{s's''}(\vqi ,\vqii)$.
Note that differences of super-indices mean the following: $\xi_{q'-q}\coleq \xi_{\vqi -\vq,s'-s}$ and $\xi_{\vq,s} = \bfa_{\vq s}+\bfa^\dagger_{-\vq s}$.
The phonon polarization vectors $\bfw_{qi}\coleq\bfw_i(\vq,s)$ satisfy the properties $\bfw_i(-\vq,s)=\bfw^*_i(\vq,s)$ and $\sum_i\bfw^*_i(\vq,s)\bfw_i(\vq,s')=\d_{ss'}$ (orthonormality).
The dimensionless phonon creation and annihilation operators satisfy the standard commutation relations
\begin{align}
 \co{\bfa_{\vq s}}{\bfa^\dagger_{\vqi s'}}&=\d_{\vq,\vqi}\d_{ss'}\qquad \forall (\vq-\vqi )\in \txt{inverse lattice vectors}
 \,, \label{eq:comm-relations}
\end{align}
and all others vanishing.
An important point to note here, is that we use, as an approximation, the isotropic Debye phonon spectrum using the effective Lam{\'e} constants of the polycrystal, i.e. ``transverse'' phonons are assumed to travel with a transverse sound speed computed from the effective polycrystalline shear modulus $\mu$.

Our Hamiltonian \eqref{eq:Hamiltonian-start} describes the interaction of phonons with (edge and screw) dislocations along the $z$-axis, moving with velocity $v$ in the $x$-direction, and depending on the two-dimensional wave vector $\vq=(q\cos\phi,q\sin\phi)$ of the dislocation.
The field of displacement gradients
due to the dislocation (in Fourier space) is denoted here by $d_{kk'}(q,\ph)$ and we will derive expressions for moving edge, screw, and mixed dislocations in Section~\ref{sec:dislocations}.
The phonon wave vectors $\vqi $, $\vqii$ lie in the first Brillouin zone and thus the dislocation wave vector satisfies $\abs{\vq}=\abs{\vqi -\vqii}\le 2q_{\txt{BZ}}$ due to momentum conservation.
For the edge of the Brillouin zone, we estimate $q_{\txt{BZ}}$ in such a way that it represents the radius of a sphere whose volume equals the unit cell volume in Fourier space, i.e.
$q_{\txt{BZ}}=\sqrt[3]{6\pi^2/\Vc }$ where $\Vc$ denotes the volume of a unit cell.

Furthermore, $\rh $ denotes the material density and the coefficients $\elA_{ijk}^{i'j'k'}$ depend on second and third order elastic constants (SOEC and TOEC), $C_{ii'jj'}$ and $C_{ii'jj'kk'}$, via~\cite{Wallace:1970,Wallace:1972}
\begin{align}
 \elA^{i'j'k'}_{ijk} &= C_{ii'jj'kk'} + C_{ii'j'k'}\d_{jk} + C_{jj'i'k'}\d_{ik} + C_{i'j'kk'}\d_{ij}
 \,. \label{eq:constantsAinC}
\end{align}
The drag coefficient (or damping/friction ``constant'') $B$ of a dislocation is defined as the proportionality coefficient of the force $F$ needed to maintain dislocation velocity $v$.
It is related to the dissipation $D$ per unit length via $D=Bv^2$, which in turn is straightforwardly derived from the probability $W_{q'q''}$ of the scattering of a phonon from state $q'$ to state $q''$ per unit time, see~\cite{Alshits:1979,Brailsford:1972}.
Multiplying $W_{q'q''}$ by the equilibrium phonon distribution function $n_{q'}=(\exp(\hbar\w_{q'}/k_BT)-1)^{-1}$ yields the number of transitions per unit time.
Taking into account that an energy $\hbar(\w_{q'}-\w_{q''})=\hbar\W_q$ is transferred for every transition, one finds for the dissipation per unit time and per unit dislocation length,
\begin{align}
 D&=\frac{4\pi}{\hbar}\sum\limits_{q',q''} \W_q |\G_{q'q''}|^2(n_{q''}-n_{q'})\d(\w_{q'}-\w_{q''}-\W_q)
 \,, \label{eq:dissipation-alshits1979}
\end{align}
where momentum conservation $\vq=\vqi \!-\vqii$ is implicit so as to avoid clutter in the notation.
The same expression can be derived from a one-loop Feynman diagram (often referred to as ``phonon wind''),
see~\cite{Alshits:1973} for details.
Since $\W_q=qv\abs{\cos\phi}$ is already linear in the dislocation velocity, $\lim\limits_{v\to0}\G_{q'q''}$ yields the small velocity limit to lowest order.
Indeed, this is what V. I. Al'shits et al. consider in Ref.~\cite{Alshits:1979}, computing $B$ for a straight line and loop dislocation for an isotropic crystal in the low velocity limit --- see also the review article~\cite{Alshits:1992}.
Ref.~\cite{Blaschke:BpaperRpt} aimed at pushing to higher velocities and hence used a $v$-dependent $\G_{q'q''}$, i.e. \eqnref{eq:Hamiltonian-start} with the displacement gradients for moving dislocations in the isotropic approximation.
For the elastic constants $\elA^{i'j'k'}_{ijk}$, experimental values for polycrystals (i.e. Lam{\'e} and Murnaghan constants) were used.

Here, our intent is to keep just the phonon spectrum isotropic and study the effect of generalizing everything else (i.e. the dislocation field and the elastic constants) to the actual single crystal symmetry, which is anisotropic in both SOEC and TOEC.

Upon introducing spherical coordinates for the phonon wave vectors and approximating the sums over those vectors by integrals over the first Brillouin zone, the drag coefficient for phonon wind $B=D/v^2$ in the continuum approximation reads~\cite{Blaschke:BpaperRpt}
\begin{align}
 B &=\frac{4\pi}{\hbar v^2}\int\limits_0^{q_{\txt{BZ}}}\!\frac{dq'\,q'^2}{(2\pi)^3}\int\limits_{-1}^1\!d\cos\th'\!\int\limits_0^{2\pi}\!d\phi'\!\int\limits_0^{2q_{\txt{BZ}}}\!\frac{dq\,q}{(2\pi)^2}\int\limits_0^{2\pi}\!d\phi\; \W_q \abs{\G_{q',q'-q}(q,\phi)}^2(n_{q'-q}-n_{q'}) \nn\\
 &\quad\times\d(\w_{q'}-\w_{q'-q}-\W_q) 
 \,, \label{eq:drag-coeff1}
\end{align}
where $\W_q=qv\abs{\cos\phi}$, aligning our coordinates such that the dislocation moves parallel to the $\hat x$ axis.
In particular,
$\vqi =q'(\sin\th'\cos\phi'\hat e_1+\sin\th'\sin\phi'\hat e_2+\cos\th'\hat e_3)$;
and then choosing these coordinates such that $\th'$ is the angle measured from the direction of $\vq$, we have
\begin{align}
(\vqi -\vq\,)^2=q^2+q'^2 -2qq'\!\cos\th'
\,. \label{eq:qnorm}
\end{align}
This means that $\hat e_3=\vq/q$, and the basis vectors $\hat e_{1,2,3}$ are hence related to the Cartesian ones via
\begin{align}
 \hat e_3&=\cos\phi\, \hat e_x+\sin\phi\, \hat e_y\,, &
 \hat e_1&=-\sin\phi\, \hat e_x+\cos\phi\, \hat e_y\,, &
 \hat e_2&=\hat e_z
 \,, \label{eq:basisvectors}
\end{align}
leading to
\begin{align}
\vq\,'\!&=q' \begin{pmatrix}
       \cos\th'\cos\ph - \sin\th'\cos\phi'\sin\ph \\
       \cos\th'\sin\ph + \sin\th'\cos\phi'\cos\ph \\
       \sin\th'\sin\phi'
       \end{pmatrix}, &
\vq &=q\begin{pmatrix}
       \cos\ph \\
       \sin\ph \\
       0
       \end{pmatrix},
 \label{eq:def-bfqi}
\end{align}
clearly consistent with \eqref{eq:qnorm}.
One of the integrals in \eqref{eq:drag-coeff1} can be evaluated using the delta function and (at least in the Debye approximation) it is convenient to eliminate the integral over $\th'$ in this way~\cite{Blaschke:BpaperRpt}.
The only $\ph'$-dependence in \eqnref{eq:drag-coeff1} comes from the kinematic factors in $\G_{q',q'\!-q}(q,\phi)$.
Therefore, the $\ph'$-integral is independent of the dispersion relation and, being of the type
$ \int_0^{2\pi}\!d\phi' \sin^m\ph'\cos^n\ph'$
with $m,n\ge0$ and $m+n\le8$, can be done explicitly.

It has been previously argued, that the dominating contribution to $B$ is due to the interaction with transverse phonons~\cite{Alshits:1992,Blaschke:BpaperRpt}, which is what we consider in the following section before subsequently generalizing further to include also longitudinal phonons.

\subsection{Interaction with transverse phonons}
\label{sec:transverse}

The Debye spectrum of transverse phonons in the isotropic limit is given by
\begin{align}
 \wt(q)&= \ct \abs{\vq}
 =\sqrt{\m/\rh }\,\abs{\vq}
 \,, \label{eq:linear-dispersion}
\end{align}
and the finite lattice spacing is taken into account indirectly by cutting off the spectrum at the Debye frequency.
One obvious shortcoming of this approximation is that it does not give a good representation of the high frequency part.
However, consistent with the continuum approximation, we consider here only
the simplest case of the Debye approximation \eqref{eq:linear-dispersion}, and leave a more thorough study of dispersion relations and their effect on the drag coefficient to future work.

Following the same steps as in Ref.~\cite{Blaschke:BpaperRpt} we introduce the variable substitution
\begin{align}
 t&=\frac{1}{2\ct^2 q{q'}}\left((\ct^2q^2+2\ct{q'}\W_q-\W_q^2\right)=\frac{1}{2 {q'}} \left(1-\bt^2\cos^2\phi\right) q +\bt\abs{\cos\phi} \,, \nn\\
 dt&=\frac{1}{2{q'}}\left(1-\bt^2\cos^2\phi\right) dq
 \,, \qquad\qquad t\in\left[\bt\abs{\cos\phi},1\right]
 \,, \label{eq:tofq}
\end{align}
where $\bt=v/\ct<1$, i.e. we have assumed a dislocation velocity below transverse sound speed of the polycrystal, a limitation of the present theory introduced by our use of the Debye spectrum.
The upper bound, $t\leq1$, is a consequence of the energy-conserving delta function in the last line of \eqnref{eq:drag-coeff1} above, which tells us that $\cos\th'=t$.
Another useful relation which follows from completeness of the polarization vectors, $\sum_s\bfw^*_i(\vqi ,s)\bfw^*_j(\vqi ,s)=\d_{ij}$, is
\begin{align}
 \sum\limits_{s=2,3}\bfw^*_i(\vqi ,s)\bfw_j(\vqi ,s)=\d_{ij}-\frac{q'_iq'_j}{q'^2}
\,,
\end{align}
where $s=2,3$ are the transverse polarizations.

Furthermore, the radial dependence of the dislocation field in the continuum limit is always $u_{i,j}(r,\th) = \tilde u_{i,j}(\th)/r$ if dislocation core effects are neglected, see Section~\ref{sec:dislocations} and Refs.~\cite{Stroh:1962,Barnett:1973,Asaro:1973,Bacon:1980}.
Hence, the same property is inherited in Fourier space if cutoffs in the $r$ integral are neglected, i.e.:
$d_{ij}(q,\phi) = \te_{ij}(\phi)/q$ for the Fourier transform of $u_{i,j}$.
Taking these considerations into account and introducing unit vectors $\hat q_i=q_i/q$, $\hat q'_i=q'_i/q'$, the dislocation drag coefficient from scattering off isotropic transverse phonons presently reads
\begin{align}
 B_\txt{tt}&=\frac{\pi\hbar}{ 4\rh ^2}\int\limits_0^{q_{\txt{BZ}}}\!\frac{dq'\,q'^4}{(2\pi)^5}\int\limits_0^{2\pi}\!d\phi \int\limits_{\bt\abs{\cos\phi}}^{1}\!\!\!dt\;  \abs{\cos\phi} \left(\frac1{e^{\frac{\hbar \ct}{\kb T}{q'}}-1}-\frac1{e^{\frac{\hbar \ct}{\kb T}{q'}(1-\bt\tilde{q}\abs{\cos\phi})}-1}\right) \nn\\
 &\quad\times\!\! \sum\limits_{\substack{i,i',j,j',k,k'\\l,m,n,l',m',n'}} \frac{\te_{kk'}(\phi)\te_{nn'}(\phi)
 }{\bt \ct^4 \left(t-\bt\abs{\cos\phi}\right)}
 \int\limits_0^{2\pi}\!d\phi' \,\hat q'_{i'}\hat q'_{l'}
 \big(\hat q'_{j'}-\tilde{q}\,\hat q_{j'}\big)\big(\hat q'_{m'}-\tilde{q}\,\hat{q}_{m'}\big)
 \nn\\
 &\quad\times \left(\d_{il}-{\hat q'_i\hat q'_l}\right)
 \left(\d_{jm}-\frac{(\hat q'_j-\tilde q\hat q_j)(\hat q'_m-\tilde q\hat q_m)}{1+\tilde q^2-2t\tilde q}\right)
 \elA_{ijk}^{i'j'k'}\elA_{lmn}^{l'm'n'}
 , \label{eq:drag-coeff-iso1}
\end{align}
where we have eliminated $q$ in favor of the dimensionless variable $t$ defined above.
Thus
\begin{align}
 \tilde{q}&\coleq\frac{q(t)}{q'}=\frac{2}{\left(1-\bt^2\cos^2\phi\right)}\left(t-\bt\abs{\cos\phi}\right)
 \,, \label{eq:def-qtilde}
\end{align}
and the delta function has already been used to integrate over $\th'$, thereby setting $\cos\th'=t$.
The latter appears in the components of the unit vector $\hat q'_i$ defined earlier,
\begin{align}
 \hat q'_i&=\begin{pmatrix}
             t\cos\ph-\sqrt{1-t^2}\sin\ph\cos\ph'\\
             t\sin\ph+\sqrt{1-t^2}\cos\ph\cos\ph'\\
             \sqrt{1-t^2}\sin\ph'
            \end{pmatrix}
 \,, \label{eq:q-prime-hat}
\end{align}
and as remarked above the integral over $\ph'$ can also be done easily.
Of the remaining three integrals, the integral over $q'$ can be evaluated in terms of Debye functions (as long as cutoffs are removed from the dislocation fields making $\te_{ij}$ independent of $q$).
These are defined as~\cite{AbramowitzStegun,Debye:1912}
\begin{align}
 D_n(x)&=\int\limits_0^x\frac{y^n}{e^y-1}dy
 =x^n\left(\frac1n-\frac{x}{2(n+1)}+\sum\limits_{k=1}^\infty\frac{B_{2k}x^{2k}}{(2k+n)(2k)!}\right)
 \,, \label{eq:Debye-fct}
\end{align}
where $\abs{x}<2\pi$, $n\ge1$, and the coefficients $B_{2k}$ are Bernoulli numbers.
In particular we have
\begin{align}
 &\int\limits_0^{q_{\txt{BZ}}}\!dq'\,q'^4\left(\frac1{e^{\frac{\hbar \ct}{\kb T}{q'}}-1}-\frac1{e^{\frac{\hbar \ct}{\kb T}{q'}(1-\bt\tilde{q}\abs{\cos\phi})}-1}\right)
 =\left(\frac{\kb T}{\hbar \ct}\right)^{\!5}\!\left(D_4\!\left(\tfrac{\hbar \ct}{\kb T}q_{\txt{BZ}}\right)-\frac{D_4\!\left(\tfrac{\hbar \ct}{\kb T}(1-\bt\tilde{q}\abs{\cos\phi})q_{\txt{BZ}}\right)}{(1-\bt\tilde{q}\abs{\cos\phi})^5}\!	\right) \nn\\
 &=\left(\frac{\kb T}{2\hbar \ct}\right)(q_{\txt{BZ}})^4\sum\limits_{k=0}^\infty\frac{B_{2k}\left(\frac{\hbar \ct}{\kb T}q_{\txt{BZ}}\right)^{\!2k}\left(1-(1-\bt\tilde{q}\abs{\cos\phi})^{2k-1}\right)}{(k+2)(2k)!}
 \,. \label{eq:Debye-fct-ourcase}
\end{align}
The series representation of these Debye functions converges only for $\hbar \ct q_{\txt{BZ}}<2\pi\kb T$, which is automatically fulfilled if $T$ is greater than the Debye temperature.
One caveat to look out for, is that the convergence of this series representation is slower as $\bt$ becomes larger (i.e. closer to 1), so that better accuracy is achieved by numerically integrating the l.h.s.
If $\bt=1$ both sides diverge, see~\cite{Blaschke:BpaperRpt}.

In deriving our results in Sec.~\ref{sec:results} below, we integrated the Debye functions numerically (rather than using the series representation \eqref{eq:Debye-fct-ourcase}) in order to achieve better accuracy.
For this we used a trapezoidal method with 400 points.
The remaining two-dimensional integral over $dt$ and $d\phi$ in \eqnref{eq:drag-coeff-iso1} above always needs to be carried out numerically, and we have done so in deriving the results of Sec.~\ref{sec:results}.
In particular, the two variables $t$ and $\phi$ were discretized with
roughly $10^5(1+\bt)$ points (i.e. with higher resolution at higher velocity) and subsequently integrated using a trapezoidal method, requiring higher resolution in $t$ than in $\phi$.
All numerical calculations described here can be reproduced with the software of Ref.~\cite{pydislocdyn} developed by the present author.

Since this strategy works for any angle-dependent Fourier transformed dislocation field $\te_{ij}(\phi)$ and any set of elastic constants $\elA^{i'j'k'}_{ijk}$, it is straightforward to generalize the purely isotropic results of Ref.~\cite{Blaschke:BpaperRpt} to a ``semi-isotropic'' calculation where $\elA^{i'j'k'}_{ijk}$ and $\te_{ij}$ are computed for the single crystal grains.

\subsection{Including longitudinal phonons}
\label{sec:longitudinal}

Even though the largest contribution to dislocation drag from phonon wind comes from the interaction with transverse phonons, the other branches cannot be completely ignored:
The combined contribution of purely longitudinal phonons ($B_\txt{ll}$) and the mixed transverse/longitudinal phonons ($B_\txt{tl}+B_\txt{lt}$) can easily increase the drag coefficient $B$ by 20\% in the low velocity regime and even more in the high velocity regime.
From \eqref{eq:drag-coeff-iso1} with \eqref{eq:Debye-fct-ourcase}, we see that $B_\txt{tt}$ scales with the fifth inverse power of transverse sound speed at low velocity and high temperature.
As we will see below, the mixed and purely longitudinal branches have two or all of those powers of $\ct$ replaced with the larger longitudinal sound speed, thus decreasing $B$ for those branches.
On the other hand, different combinations of elastic constants within $A_{ijk}^{i'j'k'}$ contribute to different branches, making the exact ratio of $B_\txt{tt}\, /\, (B_\txt{tt}+B_\txt{tl}+B_\txt{lt}+B_\txt{ll})$ material dependent (see also Fig.~\ref{fig:drag-contour-Cu-ratio} in the appendix).

The Debye spectrum of longitudinal phonons in the isotropic limit is given by $\wl(q)= \cl \abs{\vq} =\sqrt{(\l+2\m)/\rh }\,\abs{\vq}$.
For dislocations interacting with purely longitudinal phonons, the results from the previous subsection can be straightforwardly used with the simple replacements $\ct\to\cl$ everywhere, i.e. $B_\txt{ll}=B_\txt{tt}\big|_{\ct\to\cl,\bt\to\bt\ct/\cl}$.

For the mixed branches (i.e. incoming transverse, outgoing longitudinal phonon and vice versa), we however need to choose a different variable substitution instead of \eqref{eq:tofq}.
The reason is the following:
If $s'\neq s''$ the energy conserving delta function takes the more general form
\begin{align}
 \d(\w_{s'}({q'})-\w_{s''}(\abs{\vqi\!-\vq})-\W_q)
 &=\frac{\abs{c_{s'}{q'}-\W_q}}{c_{s''}^2 q{q'}}\,\d\!\left(\!\cos\th'-\!t(\qt,\phi)\right)
 \Th\left(1-\abs*{t(\qt,\phi)}\right)
 \,,\nn\\
 t(\qt,\phi)&=
\frac{\qt}{2}\left(1-\frac{v^2}{c_{s''}^2}\cos^2\phi\right)+\left(1-\frac{c_{s'}^2}{c_{s''}^2}\right)\frac{1}{2\qt} +\frac{c_{s'}v}{c_{s''}^2}\abs{\cos\phi}
 \,, \label{eq:deltafct-mixed}
\end{align}
where $\Th(x)$ is the step function following from $\cos\th'\in[-1,1]$,
and $t$ reduces to the expression linear in $\qt=q/q'$ given by \eqref{eq:tofq} only for $s'=s''$ (resp. a similar expression with $\ct\to\cl$ for the longitudinal case).
Since in general, $t(\qt,\phi)$ is a non-linear function of $\qt$, it is better to use the latter as an integration variable for the mixed transverse/longitudinal contributions to $B$.
Note that the integration range for $\qt$ is limited to finite intervals by the following conditions:
\begin{align}
 & -1\leq t(\qt)\leq 1\,, &
 & 0\leq \w_{s''}(\qt,\phi)\leq \w_\txt{BZ}\,, &
 & 0\leq\abs{\vqi\!-\vq}\leq q_\txt{BZ}
 \,. \label{eq:qt-conditions}
\end{align}
Hence $\qt_\txt{min}$, $\qt_\txt{max}$ become functions of the angle $\phi$, with finite smallest/largest values over all angles.
Within a numerical integration scheme, these conditions are hence easily implemented.

Taking these considerations into account as well as the completeness relations for the phonon polarizations, we have
\begin{subequations}
\begin{align}
 B_\txt{tl}&=\frac{\pi\hbar}{ 4\rh ^2}\int\limits_0^{q_{\txt{BZ}}}\!\frac{dq'\,q'^4}{(2\pi)^5}\int\limits_0^{2\pi}\!d\phi
 \!\!\int\limits_{\qt_\txt{min}}^{\qt_\txt{max}}\!\!\!d\qt\;  \abs{\cos\phi} \left(\frac1{e^{\frac{\hbar \ct}{\kb T}{q'}}-1}-\frac1{e^{\frac{\hbar \ct}{\kb T}{q'}(1-\bt\tilde{q}\abs{\cos\phi})}-1}\right)
 \!\sum\limits_{\substack{i,i',j,j',k,k'\\l,m,n,l',m',n'}}\!\!\! \frac{\te_{kk'}(\phi)\te_{nn'}(\phi)
 }{\bt \ct^2 \cl^2 \qt} \nn\\
 &\times\!\! 
 \int\limits_0^{2\pi}\!d\phi' \,\hat q'_{i'}\hat q'_{l'}
 \big(\hat q'_{j'}-\tilde{q}\,\hat q_{j'}\big)\big(\hat q'_{m'}-\tilde{q}\,\hat{q}_{m'}\big)
  \left(\d_{il}-{\hat q'_i\hat q'_l}\right)
 \left(\frac{(\hat q'_j-\tilde q\hat q_j)(\hat q'_m-\tilde q\hat q_m)}{1+\tilde q^2-2\qt\, t(\qt,\phi)}\right)
 \elA_{ijk}^{i'j'k'}\elA_{lmn}^{l'm'n'}
 , \label{eq:drag-coeff-mixtl}
\end{align}
\begin{align}
 B_\txt{lt}&=\frac{\pi\hbar}{ 4\rh ^2}\int\limits_0^{q_{\txt{BZ}}}\!\frac{dq'\,q'^4}{(2\pi)^5}\int\limits_0^{2\pi}\!d\phi
 \!\!\int\limits_{\qt_\txt{min}}^{\qt_\txt{max}}\!\!\!d\qt\;  \abs{\cos\phi} \left(\frac1{e^{\frac{\hbar \cl}{\kb T}{q'}}-1}-\frac1{e^{\frac{\hbar \cl}{\kb T}{q'}\left(1-\frac{\ct}{\cl}\bt\tilde{q}\abs{\cos\phi}\right)}-1}\right)
 \!\sum\limits_{\substack{i,i',j,j',k,k'\\l,m,n,l',m',n'}}\!\!\! \frac{\te_{kk'}(\phi)\te_{nn'}(\phi)
 }{\bt \ct^3 \cl \qt}
  \nn\\
 &\quad\times\!\! \int\limits_0^{2\pi}\!d\phi' \,\hat q'_{i'}\hat q'_{l'}
 \big(\hat q'_{j'}-\tilde{q}\,\hat q_{j'}\big)\big(\hat q'_{m'}-\tilde{q}\,\hat{q}_{m'}\big)
 {\hat q'_i\hat q'_l}
 \left(\d_{jm}-\frac{(\hat q'_j-\tilde q\hat q_j)(\hat q'_m-\tilde q\hat q_m)}{1+\tilde q^2-2\qt\, t(\qt,\phi)}\right)
 \elA_{ijk}^{i'j'k'}\elA_{lmn}^{l'm'n'}
 , \label{eq:drag-coeff-mixlt}
\end{align}
\end{subequations}
for the two mixed branches, where $\qt_\txt{min}(\phi)$, $\qt_\txt{max}(\phi)$ are determined by \eqref{eq:qt-conditions}.
Similar to the previous section the two variables $\qt$ and phi were discretized with
roughly $10^5(1+\bt)$ points (i.e. with higher resolution at higher velocity) and subsequently integrated using a trapezoidal method, requiring higher resolution in $\qt$ than in $\phi$.

\subsection{Dislocations}
\label{sec:dislocations}

We begin by briefly summarizing the technique of deriving the dislocation field, see e.g.~\cite{Blaschke:2017lten} and references therein.
The displacement gradient field follows from solving the equations of motion (e.o.m.) and the (leading order) stress-strain relations known as Hooke's law:
\begin{align}
 \pa_i\s_{ij}&=\rh\ddot u_j\,, &
 \s_{ij}&=C_{ijkl}\e_{kl}=C_{ijkl}u_{k,l}
 \,, \label{eq:Hooke}
\end{align}
where we have introduced the notation
$u_{k,l}\coleq\pa_l u_k$ for the gradient of the displacement field $u_k$, and $\ddot u_j\coleq \frac{\pa^2 u_j}{\pa t^2}$ for the time derivatives.
$\e_{kl}\coleq\frac12\left(u_{k,l}+u_{l,k}\right)$ denotes the infinitesimal strain tensor,
and the last equality follows from Voigt symmetry of the elastic constants.
For constant velocity the displacement field depends only on the combination $(\vec{x}-\vec{v}t)$, i.e. $u_k(x_i, t)=u_k(x_i-v_i t)$,
and thus its time derivative can be expressed in terms of its gradient: $\dot u_i=-v_j u_{i,j}$.
In this case, the e.o.m. \eqref{eq:Hooke} simplifies to
\begin{align}
0&=\pa_i\s_{ij}-\rho\ddot u_j=\left(C_{ijkl}-\rho v_iv_l \d_{jk}\right)u_{k,il}
 \,, \label{eq:eom1}
\end{align}
and it is common to define ``effective'' elastic constants $\hat C_{ijkl}\coleq\left(C_{ijkl}-\rho v_iv_l \d_{jk}\right)$, see~\cite{Bacon:1980}.

A. N. Stroh~\cite{Stroh:1962} described a method to compute solutions for infinite, straight dislocations $\vec{u}$ based on an ansatz
$ u_k=\frac{D A_k}{2\pi i}\ln\left(m_j x_j+p n_j x_j\right)$
where the perpendicular unit vectors $\vec{m}$ and $\vec{n}$ are normal to the sense vector $\vec{t}$ of the dislocation, i.e. $\vec{t}=\vec{m}\times\vec{n}$.
With this ansatz the e.o.m. \eqref{eq:eom1} is turned into an eigenvalue problem in terms of the unknown coefficients $A_k$ and $p$
and an overall factor $D$ that is determined by the boundary conditions.
Due to Voigt symmetry, the eigenvalue problem can subsequently be formulated in terms of a 6-dimensional vector $\vec\zeta$ and associated $6\times6$ matrix $\mat N$ comprised of four $3\times3$ blocks,
i.e. $\mat N\cdot\vec\zeta=p\vec\zeta$;
see~\cite[pp.~467--473]{Hirth:1982} for details on this ``sextic formalism''.

Finally, this eigenvalue problem can be reformulated in terms of a set of definite integrals~\cite{Barnett:1973,Asaro:1973},
mainly because  the unit vectors $\vec{m}$, $\vec{n}$ are defined only 
up to an arbitrary angle $\th$.
Averaging over this angle (in the notation of ref.~\cite{Hirth:1982}) yields a solution for $u_{j,k}$ in terms of the matrix
\begin{align}
 \langle \mat N\rangle&=\frac1{2\pi}\int_0^{2\pi}\mat N d\theta
 =\begin{pmatrix}
 \mat S & \mat Q \\
 \mat B & \mat S^T
 \end{pmatrix}
\,,
\end{align}
where
\begin{align}
 \mat S&=-\frac1{2\pi}\int_0^{2\pi}(nn)^{-1}(nm)\,d\theta  \,, &
\mat Q&=-\frac1{2\pi}\int_0^{2\pi}(nn)^{-1}d\theta  \,, \nn\\
\mat B&=-\frac1{2\pi}\int_0^{2\pi}\left[(mn)(nn)^{-1}(nm)-(mm)\right]d\theta
\,, \label{eq:thematrix}
\end{align}
and we have employed the shorthand notation $(ab)_{jk}\coleq a_i \hat C_{ijkl} b_l$.
These tensors depend on the elastic constants, the (constant) dislocation velocity, and material density, cf. \eqref{eq:eom1}, \eqref{eq:thematrix}.
In general, the integrals over $\th$ need to be done numerically.
Upon selecting a coordinate system such that $m_i x_i=r$, $n_ix_i=0$, the displacement gradient computes to~\cite[p.~476]{Hirth:1982}:
\begin{align}
 u_{j,k}(r,\theta)&=\frac{\tilde{u}_{j,k}(\theta)}{r}\,, \nn\\
 \tilde{u}_{j,k}(\theta)&=-\frac{b_l}{2\pi}\left\{m_k S_{jl}-n_k\left[(nn)^{-1}(nm)\right]_{ji}S_{il}-n_k(nn)^{-1}_{ji}B_{il}\right\}
 \,.\label{eq:ukl-sol}
\end{align}
Notice that the angle dependence of $\tilde{u}_{j,k}(\theta)$ resides within the unit vectors $\vec{m}$, $\vec{n}$.
The dislocation displacement gradients may subsequently be algebraically assembled according to \eqnref{eq:ukl-sol}.
In the isotropic limit $\mat S$, $\mat B$ can be calculated analytically~\cite{Bacon:1980}, and hence the moving edge and screw dislocation solutions of Eshelby are recovered~\cite{Eshelby:1949,Weertman:1980,Blaschke:2017lten}.

\begin{figure*}[h!t!b]
 \centering
 \pdfinsert{{dsq_ISO_1.5708}}{%
 In the isotropic limit, the gradient of the displacement field squared (i.e. $u_{i,j}u_{i,j}/2$) of an edge dislocation (assuming $\cl\approx2\ct$), shown here for velocities $\bt=0.001$, $\bt=0.5$, and $\bt=0.9$ where $\bt=v/\ct$,
 leads to an enhanced gradient or larger core region with increasing velocity where the assumption of linear elasticity, $u_{i,j}\ll 1$, breaks down.}
 \label{fig:contour-dsq_edge_iso}
\end{figure*}

\begin{figure*}[h!t!b]
 \centering
 \pdfinsert{{dsq_Fe_1.5708}}{%
 The gradient of the displacement field squared (i.e. $u_{i,j}u_{i,j}/2$) of an edge dislocation is shown here for (bcc) iron for velocities $\bt=0.001$, $\bt=0.5$, and $\bt=0.8$.
 In the anisotropic case illustrated here, divergences are moved to a different critical velocity $\bt^\txt{crit.}\neq1$ compared to the isotropic limit.}
 \label{fig:contour-dsq_pi2_Fe}
\end{figure*}

The slip systems we will consider are:
\begin{align}
 \vec{b}^\txt{fcc}&=\frac{b}{\sqrt{2}}\left(1,1,0\right)
 \,, &
 \vec{b}^\txt{bcc}&=\frac{b}{\sqrt{3}}\left(1,-1,1\right)
 \,, &
 \vec{b}^\txt{hcp}&=\left(-1,0,0\right)
 \,, &
 \vec{b}^\txt{tetr}&=\left(0,0,-1\right)
 \,, \nn\\
 \vec{n}_0^\txt{fcc}&=\frac{1}{\sqrt{3}}\left(-1,1,-1\right)
 \,, &
 \vec{n}_0^\txt{bcc}&=\frac{1}{\sqrt{2}}\left(1,1,0\right)
 \,, &
 \vec{n}_0^\txt{hcp}&=\left(0,0,1\right)
 \,, &
 \vec{n}_0^\txt{tetr}&=\left(0,1,0\right)
 \,, \label{eq:slipplanes}
\end{align}
i.e. these (as well as several equivalent ones due to the crystal symmetry) have the shortest Burgers vectors and are the most common ones, see~\cite[Sec. 9]{Hirth:1982} as well as~\cite{Frank:1958,Duzgun:1993} and references therein.
For the cases of close-packed hexagonal (hcp) and tetragonal crystals, we assume the basal plane is normal to the third axis in Cartesian crystal coordinates.

Following Ref.~\cite{Blaschke:2017lten} we then construct:
\begin{align}
 \vec{t}(\vth) &= \frac1b\left[\vec{b}\cos\vth+\vec{b}\times\vec{n}_0\sin\vth\right] \,, &
 \vec{v}(\vth)&=v\vec{m}_0(\vth)=v\vec{n}_0\times \vec{t}(\vth) \,,
 \nn\\
 \vec{m}(\vth,\ph)&=\vec{m}_0(\vth)\cos\ph + \vec{n}_0\sin\ph\,, &
 \vec{n}(\vth,\ph)&=\vec{n}_0\cos\ph - \vec{m}_0(\vth)\sin\ph
 \,, \label{eq:geometry}
\end{align}
where by construction $\vth$ is the angle between the Burgers vector and the dislocation sense vector $\vec{t}$, i.e. $\vth=0$ parametrizes pure screw and $\vth=\pi/2$ is pure edge.

The limitations of this approximation are illustrated in Figures~\ref{fig:contour-dsq_edge_iso} and \ref{fig:contour-dsq_pi2_Fe},
i.e. we expect a breakdown of the present theory at high velocities, typically close to the transverse sound speed of the polycrystal (which we compute from the measured effective polycrystalline shear modulus $\mu$ even in the semi-isotropic case).
The reason is that a core region where the assumption of small strains allowing the use of linear elasticity is invalid, becomes larger with increasing velocity.

In fact, as explained in Ref.~\cite{Blaschke:2017lten}, at some critical velocity which is related to the sound speeds of the crystal (and in the isotropic limit coincides with transverse/longitudinal sound speed), there is an angle $\th$ where
\begin{align}
 \det\left(nn\right)
 =\det\left(\vec{n}\cdot\mat{C}\cdot\vec{n}-\rho \left(\vec{n}\cdot\vec{v}\right)^2\id\right)
 &=0
 \,, \label{eq:divergenceinu}
\end{align}
leading to a divergence in $\tilde{u}_{j,k}(\theta)$.
Additionally, this divergence can lead to instabilities for certain types of dislocations (pure edge in the isotropic limit), driving dislocations to changing their shape~\cite{Blaschke:2017lten}.
Therefore, we will limit our discussion in Section~\ref{sec:results} to velocities $v\le0.9\ct$.

From inspecting Figures~\ref{fig:contour-dsq_edge_iso} and \ref{fig:contour-dsq_pi2_Fe} we see that depending on the degree of anisotropy and the crystal symmetry, the shapes of the contours where $u_{i,j}u_{i,j}/2\ll1$ is not fulfilled can change significantly.
But the important conclusion to keep in mind is that in all cases the area encompassing the core where $u_{i,j}u_{i,j}/2\ge1$ increases with velocity, indicating a breakdown of linear elasticity.
This also means that studying the dislocation core and its influence on dislocation drag becomes increasingly important as dislocations approach the critical velocity.
Indeed, MD simulations see significant size changes in the dislocation core at high velocities~\cite{Jin:2008}, see also~\cite{Pellegrini:2014} for a theoretical discussion in the isotropic limit in 2D.
A more realistic model of the dislocation core is however beyond the scope of the present paper and we leave it to future work.

\subsubsection*{Fourier transform of the moving dislocation deformation field}

In polar coordinates at time $t=0$, we have $(x-vt)\to r\cos\th$ and $y\to r\sin\th$ leading to $u_{i,j}(r,\th)$.
Neglecting the finite core size of the dislocation ($r_0\to0$), we may compute its Fourier transform according to
\begin{align}
 d_{ij}(q,\phi)
 = \frac{\te_{ij}(\ph)}{q}
 &=\int_0^{2\pi}d\th \;\tilde{u}_{i,j}(\th)\int_{0}^\infty dr\,e^{-iqr\cos(\th-\ph)}
 \,,
\end{align}
with $\tilde{u}_{i,j}(\th)$ computed from \eqnref{eq:ukl-sol}.
These integrations, or at least the one over $\theta$, need to be done numerically in the anisotropic case, which forces us to regularize the integral at some finite value $r_\txt{max}<\infty$, but chosen large enough to not have a noticeable effect on the result.
In particular, $r_\txt{max}=250\pi/q_\txt{BZ}$ worked well when analytically integrating $r$ and subsequently integrating $\theta$ numerically using a trapezoidal method with 3000 points.
The same angle resolution was used in the computation of $\tilde u_{i,j}$ according to Eqns.~\eqref{eq:thematrix} and \eqref{eq:ukl-sol}.
In general of course, if cutoffs are introduced $\te$ would remain $q$-dependent.

\section{Results}
\label{sec:results}

We may now compute the drag coefficient in the semi-isotropic approximation for polycrystalline metals, focusing in particular on those whose single crystal grains are of fcc, bcc, hcp and tetragonal symmetry.
Where we have enough data (such as measured Murnaghan constants), we compare the results for the drag coefficient computed using the single crystal geometry for the dislocation field and elastic constants to the purely isotropic calculation.

\begin{table*}[h!t!b]
{\renewcommand{\arraystretch}{1.1}
\centering
 \begin{tabular}{c|c|c|c|c|c|c}
          & Al\,(fcc) & Cu\,(fcc) & Fe\,(bcc) & Nb\,(bcc) & Zn\,(hcp) & Sn\,(tetr.) \\\hline
 $a$[\r{A}] & 4.05 & 3.61 & 2.87 & 3.30 & 2.67 & 5.83 \\
 $c$[\r{A}] & - & - & - & - & 4.95 & 3.18 \\
 $\rh$[kg/m$^3$]     & 2700 & 8960 & 7870 & 8570 & 7134 & 7287 \\\hline
 $\lambda$[GPa] & 58.1 & 105.5 & 115.5 & 144.5 & 43.1 & 45.9 \\
 $\mu$[GPa] & 26.1 & 48.3 & 81.6 & 37.5 & 43.4 & 18.4 \\\hline
 $\ml$[GPa] & $-143\pm13$ & $-160\pm70$ & $-170\pm40$ & $-610\pm80$ & - & - \\
 $\mm$[GPa] & $-297\pm6$ & $-620\pm10$ & $-770\pm10$ & $-220\pm30$ & - & - \\
 $\mn$[GPa] & $-345\pm4$ & $-1590\pm20$ & $-1520\pm10$ & $-300\pm20$ & - & - 
 \end{tabular}
 \caption{We list various experimental values used in the computation of the drag coefficient for some polycrystalline metals whose single crystal symmetries are fcc or bcc:
 The lattice parameters $a$, $c$ and densities $\rh$ were taken from Ref.~\cite[Sec. 12]{Haynes:2017}.
 The polycrystalline effective Lam\'e constants were taken from Refs.~\cite[p.~10]{Hertzberg:2012} and~\cite{Kaye:2004}.
 The Murnaghan constants for Cu and Fe were taken from~\cite{Seeger:1960}, those for Al were taken from Reddy 1976 as reported by Wasserb{\"a}ch in Ref.~\cite{Wasserbaech:1990}, and those for Nb were finally taken from~\cite{Graham:1968}.
 Uncertainties, as given in those references, are listed as well.
 }
 \label{tab:values-metals}
}
\end{table*}

\begin{table*}[h!t!b]
{\renewcommand{\arraystretch}{1.1}
\centering
 \begin{tabular}{c|c|c|c|c|c|c}
          & Al\,(fcc) & Cu\,(fcc) & Fe\,(bcc) & Nb\,(bcc) & Zn\,(hcp) & Sn\,(tetr.) \\\hline
 $c_{11}$[GPa]\!\! & $106.75\pm0.05$ & 168.3 & $226\pm2$ & 246.5 & 163.68 & 75.29 \\
 $c_{12}$[GPa]\!\! & $60.41\pm0.08$ & 121.2 & $140\pm8$ & 134.5 & 36.40 & 61.56 \\
 $c_{44}$[GPa]\!\! & $28.34\pm0.04$ & 75.7 & $116\pm1$ & 28.73 & 38.79 & 21.93 \\
 $c_{13}$[GPa]\!\! & - & - & - & - & 53.00 & 44.00 \\
 $c_{33}$[GPa]\!\! & - & - & - & - & 63.47 & 95.52 \\
 $c_{66}$[GPa]\!\! & - & - & - & - & - & 23.36 \\\hline
 $c_{111}$[GPa]\!\! & $-1076$ & $-1271\pm22$ & $-2720$ & $-2564\pm25$ & $-1760\pm150$ & $-410\pm15$ \\
 $c_{112}$[GPa]\!\! & $-315\pm10$ & $-814\pm9$ & $-608$ & $-1140\pm25$ & $-440\pm110$ & $-583\pm12$ \\
 $c_{123}$[GPa]\!\! & $36\pm15$ & $-50\pm18$ & $-578$ & $-467\pm25$ & $-210\pm60$ & $128\pm27$ \\
 $c_{144}$[GPa]\!\! & $-23\pm5$ & $-3\pm9$ & $-836$ & $-343\pm10$ & $-10\pm10$ & $-162\pm10$ \\
 $c_{166}$[GPa]\!\! & $-340\pm10$ & $-780\pm5$ & $-530$ & $-167.7\pm5$ & - & $-191\pm7$ \\
 $c_{456}$[GPa]\!\! & $-30\pm30$ & $-95\pm87$ & $-720$ & $136.6\pm5$ & - & $-52\pm7$
 \\\hline
 $c_{113}$[GPA]\!\! & - & - & - & - & $-270\pm30$ & $-467\pm12$ \\
 $c_{133}$[GPa]\!\! & - & - & - & - & $-350\pm10$ & $-186\pm4$ \\
 $c_{155}$[GPa]\!\! & - & - & - & - & $250\pm50$ & $-177\pm10$ \\
 $c_{222}$[GPa]\!\! & - & - & - & - & $-2410\pm260$ & - \\
 $c_{333}$[GPa]\!\! & - & - & - & - & $-720\pm20$ & $-1427\pm9$ \\
 $c_{344}$[GPa]\!\! & - & - & - & - & $-440\pm40$ & $-212\pm11$ \\
 $c_{366}$[GPa]\!\! & - & - & - & - & - & $-78\pm14$
 \\\hline
 $A$ & 1.22 & 3.28 & 2.70 & 0.51 & - & -
 \end{tabular}
 \caption{We list the experimental values for the single crystal elastic constants of various metals used in the semi-isotropic computation of the drag coefficient.
 SOEC are taken from the CRC handbook~\cite[Sec. 12]{Haynes:2017}, the original references being~\cite{Thomas:1968} (Al), \cite{Epstein:1965} (Cu), \cite{Leese:1968} (Fe), \cite{Bolef:1961} (Nb), \cite{Alers:1958} (Zn), and \cite{House:1960} (Sn).
 TOEC are taken from \cite{Thomas:1968} (Al), \cite{Hiki:1966} (Cu), \cite{Powell:1984} (Fe), \cite{Graham:1968} (Nb), \cite{Swartz:1970} (Zn), and \cite{Swartz:1972} (Sn).
 Uncertainties, as given in the original references, are listed as well.
 In the last line we list the Zener anisotropy for cubic metals computed from the second order elastic constants according to $A\coleq 2c_{44}/(c_{11}-c_{12})$.
 }
 \label{tab:values-metals_cubic}
}
\end{table*}

\subsection{Cubic metals}

The effective isotropic elastic constants that we use as (experimental) input data are assembled in Table~\ref{tab:values-metals}.
For the unit cell volume of cubic metals we use $\Vc =a^3$, and for the (length of the) Burgers vector $b=a/\sqrt{2}$ for fcc metals and $b=a\sqrt{3}/2$ for bcc metals (see Refs.~\cite{Frank:1958} and~\cite[Sec. 9]{Hirth:1982}).
The single crystal constants are assembled in Table~\ref{tab:values-metals_cubic}.
The selection of metals presented here (Al, Cu, Fe, Nb), were guided by the availability of experimental effective TOEC for polycrystals in the literature.
We refrain here from computationally averaging over single crystal constants, as there is no good averaging scheme for the TOEC, see~\cite{Blaschke:2017Poly} and references therein.

\begin{figure*}[h!t!b]
 \centering
 \pdfinsert{{B_room_semiiso_compare_0.0000}}{%
On the left, we show the drag coefficient from phonon wind for screw dislocations in various metals in the isotropic approximation with zero dislocation core size (i.e. no cutoff), using isotropic SOEC and TOEC from Table~\ref{tab:values-metals}.
On the right, we show the drag coefficient for screw dislocations in the semi-isotropic approximation, using cubic SOEC and TOEC from Table~\ref{tab:values-metals_cubic},
but the polycrystal value of the shear modulus for the transverse phonons and the according effective transverse sounds speed $\ct$.}
 \label{fig:dragiso-screw_compare}
\end{figure*}

\begin{figure*}[h!t!b]
 \centering
 \pdfinsert{{B_room_semiiso_compare_1.5708}}{%
On the left, we show the drag coefficient from phonon wind for edge dislocations in various metals in the isotropic approximation with isotropic elastic constants.
On the right, we show the drag coefficient for edge dislocations in the semi-isotropic approximation, using cubic elastic constants.}
 \label{fig:dragiso-edge_compare}
\end{figure*}

After computing the dislocation displacement gradients (numerically) along the lines of Section~\ref{sec:dislocations}, one must rotate them in order to align the dislocation sense vector with the $\hat z$-axis, and the dislocation velocity with the $\hat x$ axis.
This step is important, since the expression for the drag coefficient, \eqnref{eq:drag-coeff-iso1}, was derived in a coordinate system where $\vec{t}\parallel\hat{z}$ and $\vec{v}\parallel\hat{x}$, whereas $u_{i,j}$ is computed in crystal coordinates initially.
Since in the derivation of $u_{i,j}$ the Burgers vector was fixed and the direction of $\vec{t}$ depended on the angle $\vth$ defining the dislocation type, this required additional rotation is also $\vth$ dependent.
The same rotation must also be applied to the elastic constants $\elA_{ijk}^{i'j'k'}$, as these are also given in crystal coordinates.
In particular, for fcc metals we must rotate around $\hat z$ by $\pi/4$, then around $\hat x$ by $-\txt{atan}(1/\sqrt{2})$, and then around $\hat y$ by $\pi/2-\vth$.
For bcc metals we must rotate around $\hat z$ by $-\pi/4$ and then around $\hat y$ by $\pi/2-\txt{atan}(1/\sqrt{2})-\vth$.
All calculations in this section and the next were done for 90 velocities in the range $0.01\le\bt\le0.9$ and 91 angles in the range $0\le\vth\le\pi/2$ (resp. 181 angles in the range $-\pi/2\le\vth\le\pi/2$ depending on the symmetry properties of the slip system).

Inspecting Figures~\ref{fig:dragiso-screw_compare}, \ref{fig:dragiso-edge_compare}, we notice two types of changes in going from the purely isotropic to the semi-isotropic calculation:
In the high velocity range, the fast growth (resp. divergence) of the drag coefficient is moved from $\ct$ to a different velocity which depends on the dislocation character angle $\vth$ and which can be computed from \eqnref{eq:divergenceinu}.
Hence, even pure screw dislocations diverge at their critical velocity.
This behavior contrasts the isotropic limit where this divergence was suppressed by the polynomial parts of the expression for $B$, cf. \eqref{eq:drag-coeff-iso1}.
These depend on $\ct$ through the (Debye) phonon spectrum, even in the semi-isotropic calculation.
This, however, may change once we generalize from the Debye to the actual phonon spectrum in the crystal, which is beyond the scope of the present work.
The second noticeable change is in the low velocity regime of $B$, where the change in the drag coefficient for pure screw and edge dislocations clearly depends on the degree of anisotropy in the second order elastic constants --- which is to be expected.
Hence, while $B$ for aluminum changes only very little in the low velocity regime, $B$ for niobium changes significantly.

Furthermore, the drag coefficient for a mixed dislocation
coincides with a linear superposition only in the isotropic case, but not in the semi-isotropic approximation.
In particular,
\begin{align}
 B_\txt{iso}(\vth) &= \cos^2\vth\; B_\txt{iso}(\vth=0)+\sin^2\vth\; B_\txt{iso}(\vth=\tfrac{\pi}2)
 \,.
\end{align}
The reason for this relation is that the displacement gradients of screw and edge dislocations decouple only in the isotropic limit,
and additionally cross terms in $B$ vanish in this limit.
In general, however, $B$ is a non-trivial function of dislocation character angle $\vth$.
To
illustrate this point, we show the explicit $\vth$ dependence of the drag coefficient for copper at several velocities for both the isotropic and the semi-isotropic limit in Figure~\ref{fig:dragiso-Cu_compare}.

\begin{figure*}[h!t!b]
 \centering
 \pdfinsert{{B_room_semiiso_compare_Cu}}{%
On the left, we show the drag coefficient from phonon wind for copper as a function of dislocation type (i.e. $\vth=0$ is pure screw, $\vth=\pi/2$ is pure edge) for various velocities.
Since edge and screw components decouple in this limit, $B_\txt{mix}=B_\txt{screw}+B_\txt{edge}$.
On the right, we show the same in the semi-isotropic approximation:
Edge and screw components no longer decouple in this limit and hence $B_\txt{mix}$ turns into a nonlinear function of $\sin^2\vth$.}
 \label{fig:dragiso-Cu_compare}
\end{figure*}

\begin{figure}[h!t!b]
 \centering
 \includegraphics[width=0.6\textwidth]{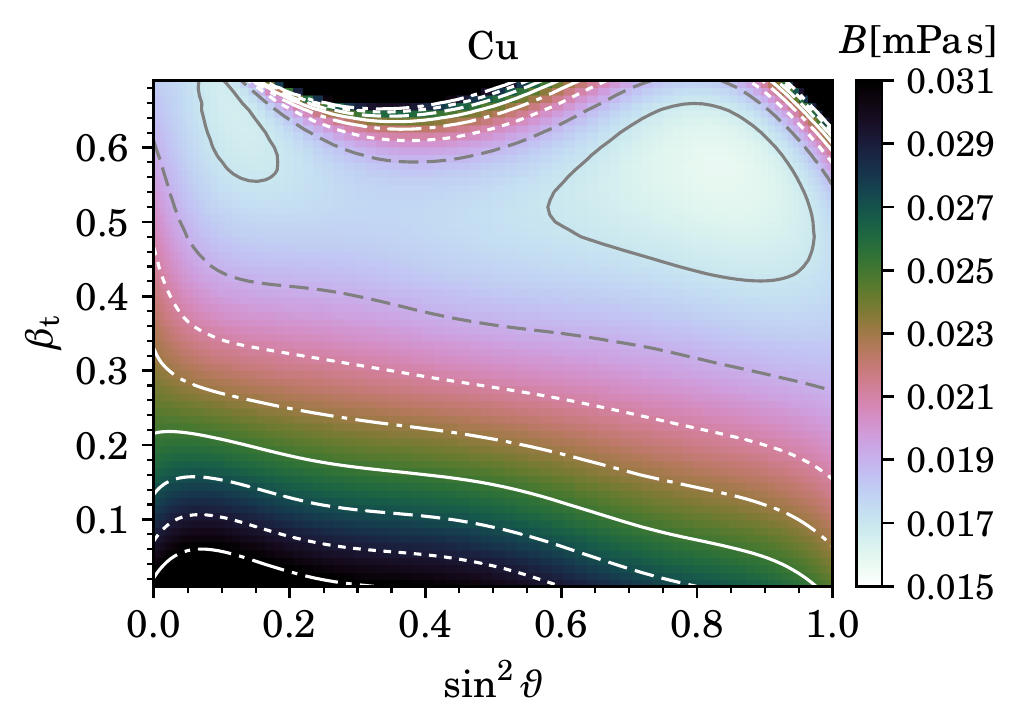}
\caption{%
We show the drag coefficient $B$ from phonon wind for copper as a function of dislocation type $\vth$ and velocity $\bt$.
$B(\bt,\vth)$ is color-encoded and given in units of mPa\,s with contour lines drawn at the values given in the color bar.}
 \label{fig:drag-contour-Cu}
\end{figure}

The contour plot in Figure~\ref{fig:drag-contour-Cu} finally captures the full dependence on velocity $\bt$ and angle $\vth$ of the drag coefficient in copper.
The velocity dependence is shown up to the smallest critical velocity $\bt\le \beta_\txt{crit}=v_\txt{crit}/\ct\approx 0.698$ for copper, i.e. the lowest velocity at which a divergence appears in the dislocation field for some angle $\vth$ ($=\pi/2$ for fcc) due to \eqnref{eq:divergenceinu}, see Ref.~\cite{Blaschke:2017lten}.
As for the character angles $\vth$ at which $B(v)$ takes its largest values, we note that we are dealing with a non-trivial interplay of crystal geometry, material constants, and kinematics:
The dislocation gradient field peaks where the determinant \eqref{eq:divergenceinu} becomes small.
This expression depends not only on the crystal geometry, but also on density, velocity and elastic constants, leading to peaks at various combinations of character angle $\vth$, velocity $v$, and polar angle $\phi$.
Finally, the kinematics within $B$ (due to energy-momentum conservation) do not allow all angles between dislocation wave vector $\vq$ and phonon wave vector $\vq'$ to contribute, potentially cutting off regions of large dislocation fields (especially at high velocity).

\begin{figure}[h!t!b]
 \centering
 \includegraphics[width=0.88\textwidth]{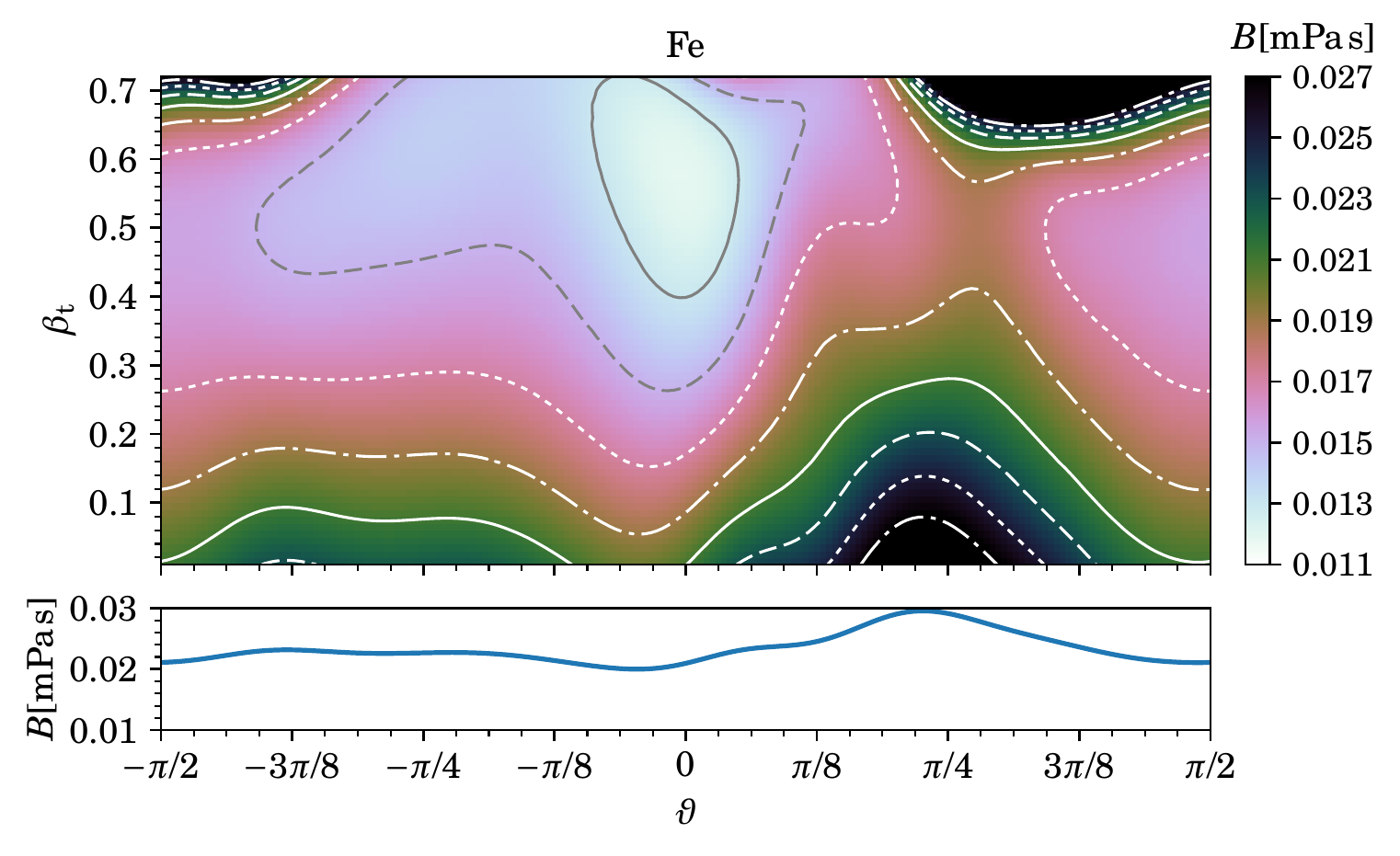}
\caption{%
We show the drag coefficient $B$ from phonon wind for iron as a function of dislocation type $\vth$ and velocity $\bt$.
$B(\bt,\vth)$ is color-encoded and given in units of mPa\,s with contour lines drawn at the values given in the color bar.
At the bottom of this figure, we show the low velocity drag coefficient $B(\vth)\big|_{\bt=0.01}$ as a function of $\vth$.}
 \label{fig:drag-contour-Fe}
\end{figure}

Also note that while the drag coefficient for fcc metals is symmetric\footnote{
To be precise, it is symmetric for a perfect dislocation (which is what we consider here), i.e. the inclusion of two different Shockley partial dislocations will likely break this symmetry.
I thank the anonymous referee for pointing out this subtlety.
}
with respect to $\vth\leftrightarrow-\vth$, this is not the case for the slip system considered here for bcc metals; only the pure edge cases $\vth=\pm\pi/2$ coincide and both dislocation field and drag coefficient are $\pi$-periodic.
This asymmetry with respect to dislocation character (or rather orientation of the dislocation) is illustrated at the example of bcc iron in Figure~\ref{fig:drag-contour-Fe}, now plotted against $\vth$ instead of $\sin^2\vth$.
Once more, the velocity dependence is shown up to the lowest critical velocity which is $\beta_\txt{crit}\approx 0.726$ for iron.

\subsection{Hexagonal and tetragonal metals}

The semi-isotropic approximation outlined above, allows us to make use of the much greater wealth of experimentally measured single crystal TOEC.
The methods described here can be straightforwardly applied to crystal symmetries other than fcc/bcc, as we demonstrate in this subsection:
The slip systems we consider for hcp and tetragonal crystals are already summarized in \eqnref{eq:slipplanes}, and in both cases they lead to dislocation fields and hence drag coefficients which are symmetric with respect to $\vth\leftrightarrow-\vth$.
Additionally, we need expressions for the length of the Burgers vector as well as for the unit cell volume which is then used to determine the edge of the Brillouin zone, as outlined earlier.
In particular, $\Vc =\frac{3\sqrt{2}}2a^2c$ for hcp metals and $\Vc =a^2c$ for tetragonal metals.
Furthermore, $b=a$ for hcp metals and $b=c$ for tetragonal metals and the slip systems we consider here.
For hcp metals we must rotate both $\te_{ij}$ and $\elA_{ijk}^{i'j'k'}$ around $\hat y$ by $-\pi/2-\vth$ and then around $\hat x$ by $\pi/2$.
For tetragonal metals we must rotate  around $\hat y$ by $\pi-\vth$.

\begin{figure}[h!t!b]
 \centering
 \includegraphics[width=\textwidth]{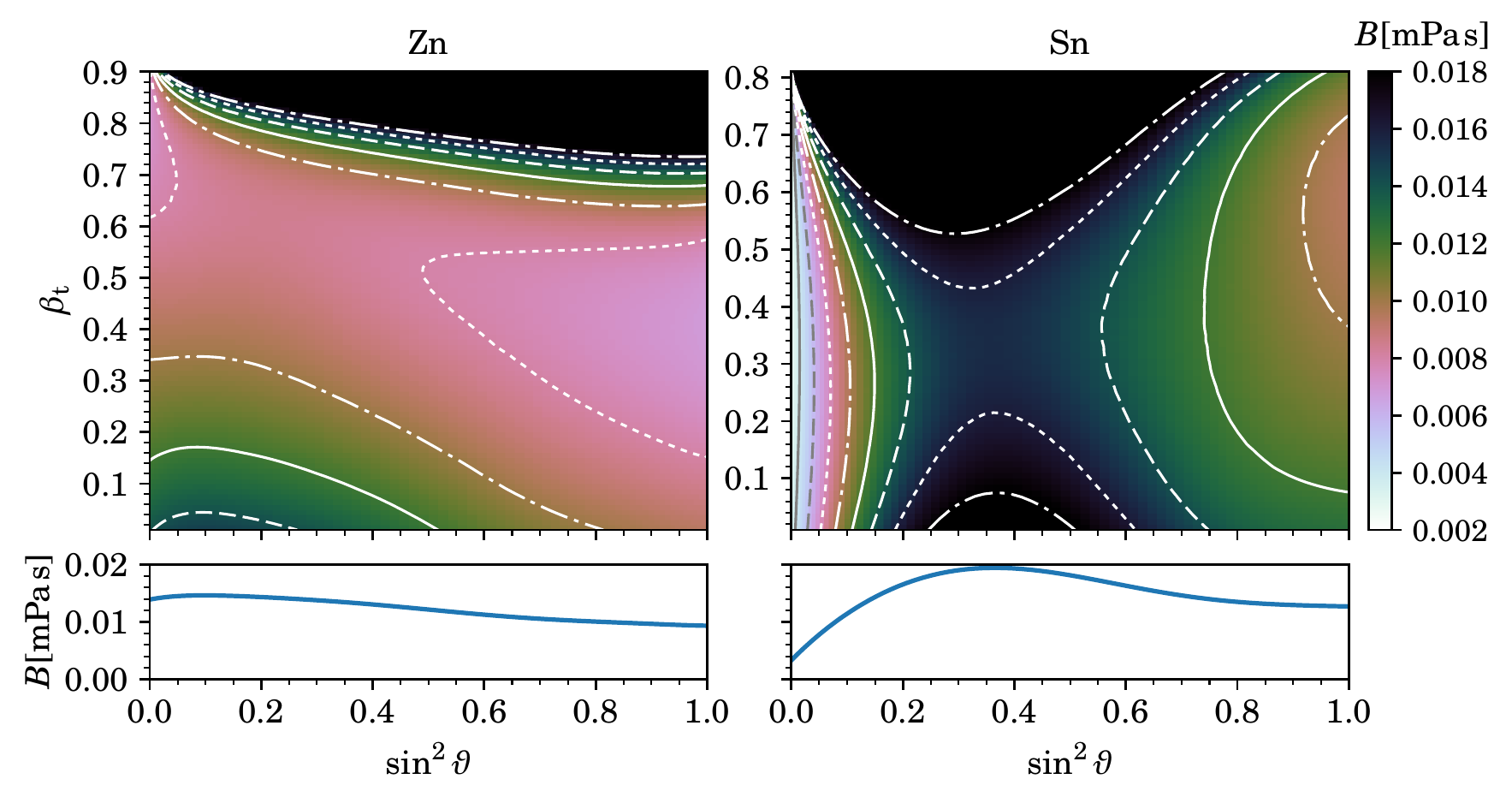}
\caption{%
We show the drag coefficient $B$ from phonon wind for hexagonal close-packed zinc (lhs) and for tetragonal white tin (rhs) as a function of dislocation type $\vth$ and velocity $\bt$.
$B(\bt,\vth)$ is color-encoded and given in units of mPa\,s (cf. color bar).
At the bottom of this figure, we show the low velocity drag coefficient $B(\vth)\big|_{\bt=0.01}$ as a function of $\vth$.}
 \label{fig:drag-contour-ZnSn}
\end{figure}

As examples for both types of crystal symmetry, we present calculations for the drag coefficient from phonon wind for zinc (hcp) and white tin (which is tetragonal and also commonly referred to as $\beta$-tin).
The experimental input values needed are listed in Tables~\ref{tab:values-metals} and \ref{tab:values-metals_cubic}.
The drag coefficient $B$ was then calculated numerically along the same lines as in the previous section.
Results are shown in Figure~\ref{fig:drag-contour-ZnSn}.
In particular, the contour plots shown in that figure
capture the full dependence on velocity $\bt$ ($\le\beta_\txt{crit}$ and $\le0.9$) and angle $\vth$ of the drag coefficient in those two metals.
The smallest critical velocities leading to divergences due to \eqref{eq:divergenceinu} are $\beta_\txt{crit}=v_\txt{crit}/\ct\approx0.943$ for zinc and $\beta_\txt{crit}\approx0.818$ for tin.
The values for $B$ at $\bt=0.01$ are shown separately at the bottom of this figure as a function of $\vth$.

\subsection{Comparing to experimental and MD simulation results}

Comparing our results at low velocity to experiments and MD simulations, we note that
\begin{itemize}
\itemsep=0pt
 \item Our drag coefficient (in both approximations discussed above) for Al (fcc aluminum) lies within the range of experimental values of $\sim0.005\,$mPas in~\cite{Hikata:1970}, $\sim0.02\,$mPas in~\cite{Gorman:1969} and $\sim0.06\,$mPas in~\cite{Parameswaran:1972}, and at the lower end of MD simulation results which range from $\sim0.007\,$mPas to $\sim0.2\,$mPas~\cite{Olmsted:2005,Yanilkin:2014,Cho:2017}, with slightly better agreement within the semi-isotropic approximation;
 see Figures~\ref{fig:dragiso-screw_compare}--\ref{fig:dragiso-edge_compare} (values for $\bt=0.01$).
 
 \item The drag coefficient (in both approximations discussed above) for Cu (fcc copper) is well within the range of experimental values of $\sim0.0079\,$mPas in~\cite{Suzuki:1964}, $\sim0.02\,$mPas in~\cite{Zaretsky:2013},
 $\sim0.065\,$mPas (for both edge and screw dislocations) in~\cite{Stern:1962},
 $\sim0.07\,$mPas in~\cite{Greenman:1967},
 and $\sim0.08\,$mPas in~\cite{Alers:1961}.
 It is above the MD simulation results of $\sim0.016\,$mPas (edge) and $\sim0.021\,$mPas (screw) reported in~\cite{Oren:2017}, and $\sim0.022\,$mPas (screw) reported in~\cite{Wang:2008};
 see Figure~\ref{fig:dragiso-Cu_compare} (values for $\bt=0.01$).
 
 \item Our drag coefficient (in both approximations discussed above) for Fe (bcc iron) is lower than the experimental values of $\sim0.34\,$mPas for edge and $\sim0.661\,$mPas for screw dislocations reported in~\cite{Urabe:1975},
 as well as the result of MD simulations of $\sim0.26\,$mPas for screw dislocations reported in~\cite{Gilbert:2011};
 see Figures~\ref{fig:dragiso-screw_compare}--\ref{fig:dragiso-edge_compare}
 and the lower part of Figure~\ref{fig:drag-contour-Fe} (values for $\bt=0.01$).




 \item The drag coefficient for Zn (hcp zinc), computed in the semi-isotropic approximation, is lower than the experimental results of $0.034\,$mPas for screw and $0.035\,$mPas for edge dislocations in the basal plane~\cite{Pope:1969};
 see lower lhs of Figure~\ref{fig:drag-contour-ZnSn} (for $\bt=0.01$).
\end{itemize}
The experimental values we compare to are typically either of mixed edge/screw type or unknown (unless we have stated explicitly otherwise above).
With increasing dislocation velocity, we have qualitative agreement of our semi-isotropic approach with simulation results, i.e. we see a viscous regime where $B(v)$ for pure screw and edge changes only little and a regime close to the critical velocity where damping is enhanced.
The only discrete lattice simulation which considered the dislocation character angle dependence the author is aware of is Ref.~\cite{Cho:2017} on aluminum.
The authors of~\cite{Cho:2017} find a stronger $\vth$-dependence for small $v$ than in this work by fitting their simulation results to Leibfrieds isotropic high temperature approximation to $B$, i.e. constant in $v$ and linear in $T$, thus ignoring both the (small but not negligible) velocity dependence in the viscous regime as well as the non-linear temperature dependence between $100$--$300$K.

\section{Conclusion and outlook}
\label{sec:discussion}

In this work, the dislocation drag coefficient $B$ from phonon-scattering (``phonon wind'') at room temperature was revisited in the continuum approximation, and the model was subsequently generalized to include anisotropic effects from the single crystal grains in a poly-crystalline metal.
The reason anisotropic effects are important is that dislocations move through single crystal grains which are much larger than a Burgers vector and may pass through grain boundaries, but they never ``see'' an isotropic medium;
Effective isotropic properties are at a macroscopic scale of the polycrystal only.

As a first step towards a fully anisotropic model, we considered here the interaction between dislocations and elastic constants of the  anisotropic single crystal grains and an isotropic Debye phonon spectrum of the polycrystal.
The Debye spectrum and several other approximations greatly simplified the theory:
In particular, we limited ourselves to monatomic lattices and considered the approximation of linear elasticity (i.e. small lattice displacements and small displacement gradients).
The continuum approximation, constant subsonic dislocation velocity as well as neglecting dislocation core effects led to further significant simplifications.
On the other hand, in order to take into account the anisotropy of the single crystal grains to some extent, we derived the dislocation gradient field in the full anisotropic theory for dislocations moving at constant velocity.
This ``semi-isotropic'' approach is considered as an intermediate step in an ongoing long-term endeavor to include all anisotropic effects and the true phonon spectrum, but this is beyond the scope of the current work.

Nonetheless, we already gained valuable insights, like the non-trivial dependence of the drag coefficient on the dislocation character angle $\vth$ (between line sense and Burgers vector) shown in Figures~\ref{fig:dragiso-Cu_compare}--\ref{fig:drag-contour-ZnSn}.
Especially, the high velocity regime (i.e. close to transverse sound speed) changes significantly if anisotropic effects are taken into account:
The dislocation displacement gradient field exhibits divergences at certain combinations of velocity and angle $\vth$, and these are different from the purely isotropic approximation, which is not able to capture the rich interplay of crystal and dislocation geometry and dislocation velocity.

Due to the lack of experimental data at high dislocation velocities, we could compare our present predictions only in the ``low'' velocity limit, by which we mean the ``viscous'' regime of about 1\% transverse sound speed.
Our results are within the range of experimental data for copper and aluminum, but lower than the experimental results for iron and zinc.
No experimental data on the drag coefficient could be found for niobium or tin.
Several reasons for this discrepancy can be envisioned:
First and foremost, we considered only the scattering with phonons within an isotropic Debye spectrum, which deviates from the true one especially in the high frequency regime.
Taking into account the full anisotropic phonon spectrum might increase $B$, and more so for some metals than others.
Dislocation core effects, on the other hand, tend to decrease the drag coefficient~\cite{Alshits:1992,Blaschke:BpaperRpt}.
Additionally, the interaction with grain boundaries may be a factor, as well as high uncertainties in the experimental determination of dislocation drag (cf. the range of experimental values for copper) as well as TOEC, which are both hard to measure accurately.
In particular, the uncertainties in the TOEC affect the accuracy of our present predictions, being one of the major uncertainty sources.
In comparing to MD simulation results at higher velocities we see qualitative agreement, although our present first principle results show far more detail in the velocity and dislocation character dependence.

In order to improve future predictions, it is worthwhile to study the full anisotropic phonon spectrum.
Additional and more accurate experimental data for TOEC as well as dislocation drag (for comparison and validation) would also be very helpful.
Further future improvements to be considered include the full temperature dependence of $B$ (work in progress), 
the inclusion of dislocation core effects,
as well as the generalization to dislocations accelerating to transonic and supersonic speeds.

\subsection*{Acknowledgements}

I thank Darby J. Luscher, Dean L. Preston, and Benjamin A. Szajewski for enlightening discussions.
I also thank the anonymous referee for valuable comments.
This work was performed under the auspices of the U.S. Department of Energy under contract DE-AC52-06NA25396.
In particular, the author is grateful for the support of the Advanced Simulation and Computing, Physics and Engineering Models Program.

\appendix
\section{Appendix: Interaction Hamiltonian for phonon wind}
\label{sec:appendix}

We briefly summarize the most essential steps necessary to derive the Hamiltonian \eqref{eq:Hamiltonian-start} and subsequently the expression for the drag coefficient from phonon wind \eqref{eq:drag-coeff1}, and we refer to Refs.~\cite{Blaschke:BpaperRpt,Alshits:1992} for further details.
Our starting point is the crystal potential, Taylor expanded in terms of the finite Murnaghan strains $\eta_{ij} = \inv2\left(U_{i,j}+U_{j,i}+U_{k,i}U_{k,j}\right)$, where $U_{i,j} = \partial_j U_i$ are gradients of the displacement field in the continuum limit:
\begin{align}
\Phi&=\Phi_0+C_{ij}\eta_{ij}+\inv{2}C_{ijkl}\eta_{ij}\eta_{kl}+\inv{3!}C_{ijklmn}\eta_{ij}\eta_{kl}\eta_{mn}+\ldots
\end{align}
Furthermore, we assume $U_i =u_i^\txt{dis}+u_i^\txt{ph}$ consists of a linear superposition of displacements due to a dislocation field $u_i^\txt{dis}$ and phonons $u_i^\txt{ph}$.
Being interested in phonons scattering off the dislocation, we then need only consider the kinetic term, bilinear in the phonons, and the interaction term proportional to $u_{i,i'}^\txt{dis} u_{j,j'}^\txt{ph} u_{k,k'}^\txt{ph}$ (i.e. an interaction vertex with incoming phonon and outgoing phonon after scattering off a dislocation).
Because $\eta_{ij}$ is quadratic in the displacement gradients, we get contributions to this interaction term depending on the second order as well as third order elastic constants, see \eqnref{eq:constantsAinC} above.
For our computations we require the Fourier transforms of these terms.
The according expression for the dislocation $u^\txt{dis}_{i,j}$ in the continuum limit is derived above in Section~\ref{sec:dislocations}, and upon choosing coordinates aligned in the $\hat z$ direction with the dislocation sense vector, its spatial dependence is only two-dimensional, i.e. $u^\txt{dis}_{i,j}(x,y)$ (or $d_{ij}(q,\phi)$ in Fourier space with polar coordinates).

The phonons are quantized in the usual way and written as
\begin{align}
u_i^\txt{ph} &= \sqrt{\frac{\hbar}{2\rho}}\sum_{q,s}\w_s(\vq)^{-1/2}\left(\bfa_{\vq s}e^{i\vq\vec{x}-i\w_st}+\bfa^\dagger_{\vq s}e^{-i\vq\vec{x}+i\w_st}\right)\bfw_{i}(\vq,s) 
\,,
\end{align}
and since the only $x$ dependence resides in the exponent, its gradient is easily derived, bringing down factors $\pm i\vq$.
The phonon creation and annihilation operators $\bfa_{\vq s}^\dagger$ and $\bfa_{\vq s}$, satisfy \eqref{eq:comm-relations} and the polarization vectors are orthonormal (cf. Section~\ref{sec:phononwind}).
Plugging this expression into $\int\!d^4x\,\Phi$ (Fourier transformed) and taking into account, apart from energy-momentum conservation, that the only time-dependence of a steady-state continuum solution to the dislocation field in Fourier space is via its ``frequency'' $e^{-i\W_q t}$, we find (among other terms which do not contribute to the problem at hand) the Hamiltonian of \eqref{eq:Hamiltonian-start}.
Finally, in order to arrive at \eqref{eq:drag-coeff1}, we additionally approximated the sum over the phonon wave vectors by integrals over the first Brillouin zone (BZ), $\sum{\vq}\to V\int_\txt{BZ}\frac{d^3q}{(2\pi)^3}$, and wrote the latter in spherical coordinates.
Hence, we approximate the BZ by a sphere of the same volume leading to the relation $(2\pi)^2/\Vc = (4/3)\pi/q_\txt{BZ}^3$ between the radius of this sphere $q_\txt{BZ}$ in Fourier space and the unit cell volume $\Vc$.

The drag coefficient $B$ is defined via $F = Bv$, i.e. it is the proportionality of the force $F$ needed to maintain velocity $v$.
Assuming that all the friction is dissipated as heat, and due to phonon wind, one deduces that $B=D/v^2$ and as described above, $D$ can be computed via \eqref{eq:dissipation-alshits1979}, subsequently leading to \eqref{eq:drag-coeff1} with the approximations outlined above.

\begin{figure}[h!t!b]
 \centering
 \includegraphics[width=\textwidth]{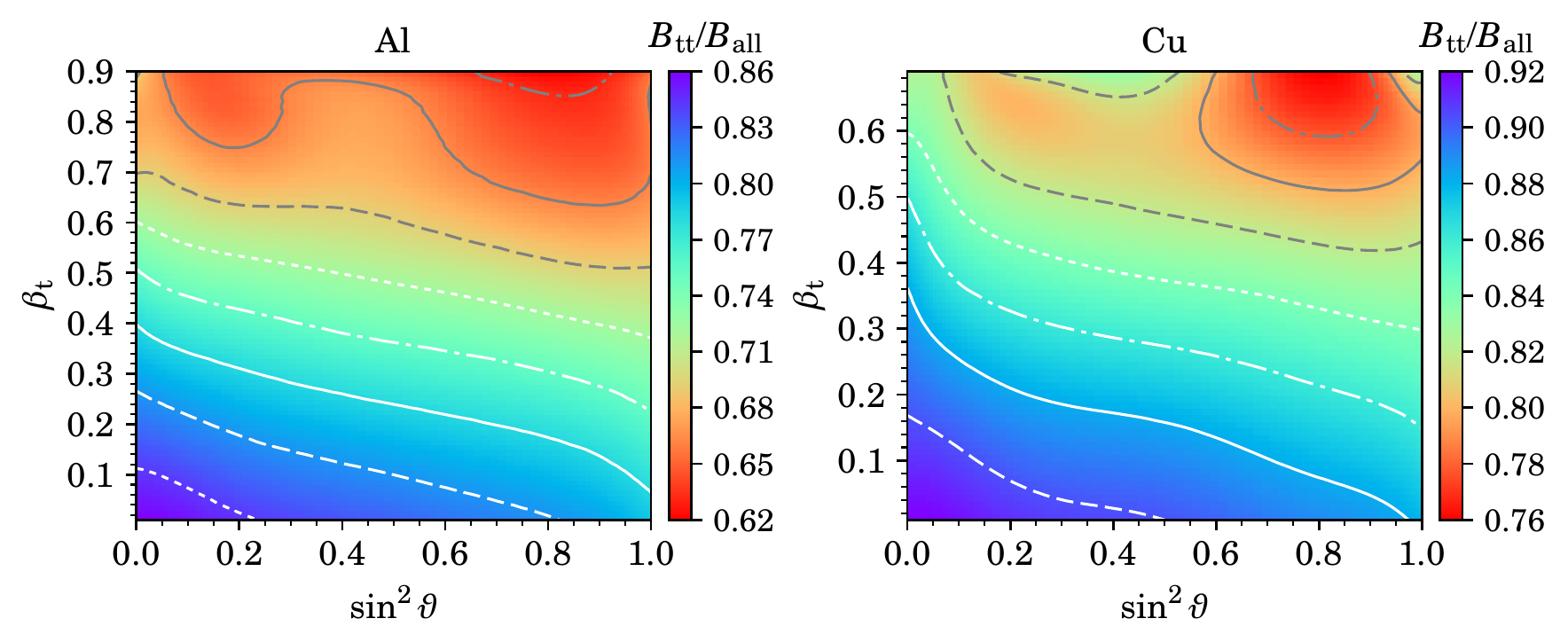}
\caption{%
We highlight the importance of transverse phonons for the drag coefficient $B$ by  showing the ratio of $B_\txt{tt}$ from scattering purely transverse phonons over $B_\txt{all}=(B_\txt{tt}+B_\txt{tl}+B_\txt{lt}+B_\txt{ll})$ at the examples of aluminum and copper as functions of dislocation type $\vth$ and velocity $\bt$.
While transverse phonons typically contribute 80\% and more to $B$ at low velocity, the other branches (in particular the mixed ones) gain additional importance at higher velocities.
}
 \label{fig:drag-contour-Cu-ratio}
\end{figure}

As noted in Section~\ref{sec:longitudinal} above, the most important contribution to $B$ is due to the scattering of purely transverse phonons in the (semi-)isotropic approximation, although the other branches cannot be neglected either.
Figure~\ref{fig:drag-contour-Cu-ratio} quantifies this statement for the examples of fcc aluminum and copper.

\bibliographystyle{utphys-custom}
\bibliography{dislocations}

\end{document}